\documentstyle[preprint,aps,prb]{revtex}
\tightenlines
\begin{document}
\title{Van Hove Singularities around the Fermi level in YBa$_{2}$Cu$_{3}$O$_{7}$:
The importance of the chains.}
\author{A. Rubio-Ponce${^*}$ and R. Baquero \\
Departamento de F\'{\i}sica, CINVESTAV,\\
A. Postal 14-740, 07000 M\'{e}xico D.F.}
\maketitle

\begin{abstract}
We have reproduced band structure calculations from the literature and have
used them to analyze in detail the energy landscape around the Fermi level, E%
$_{F}$. We found three Van Hove singularities, two below (-230, -54 meV) and
one above the Fermi level (+27 meV). We have studied the composition of each
one of them and found that states comming from the chain do contribute in a
very importan way. The contribution from the planes are indeed important
and, therefore, we find that a 2D description includes the most important
contributions. Nevertheless, the contribution from states out of the planes
(the chains and the apical oxygen) is by no means negligable. We find that
it is possible that in YBaCuO part of the condensate lies in the chains, a
fact that would agree with some recent evidences concerning PrBaCuO. Our
general conclusion is that the 2D description of 123-compounds might be
insuficient to explain all the experimental details and that a 3D
description seems compulsory to fully account for the phenomenom of
superconductivity in YBaCuO.


\noindent {\it Author to whom the corresponce should be sent to:}

rbaquero@fis.cinvestav.mx
\end{abstract}

\section{Introduction\ {\protect\small \ }}

In a recent paper, Quesada {\it et al.}\cite{davalb98} have reviewed the
thermodynamics of YBa$_{2}$Cu$_{3}$O$_{7}$ within the van Hove scenario. We
find that it explains a number of experiments, as Bok and Bouvier have
stated recently \cite{bokbouvier}. Nevertheless, some of the results deviate
from the experimental ones, sometimes considerably. The inclusion of other
scenarios as active in the superconducting state could bring more light into
the problem, first, because the three-dimensional dispersion (c-axis) can be
important as Pickett {\it et al.} \cite{deweert89} have suggested, and
secondly, because the condensate might lie (at least in part) in the chains.
Indeed, Cucolo {\it et al. }\cite{cucolo}{\it \ }made a simulation of their
experimental results for YBa$_{2}$Cu$_{3}$O$_{7}$ (tunneling, specific heat
and ultrasonic attenuation) where they have included the CuO$_{2}$ planes,
the b-axis CuO chains and the c-axis as active scenarios. Their description
of each scenario is particularly simple and still the three experiment are
brought into agreement. For superconducting PrBa$_{2}$Cu$_{3}$O$_{7}$, Dow
and Blackstead \cite{dow99} reported, very recently, experimental results
and interpreted them as demonstrating that the condensate lies in the chains.

In this paper, we show that the CuO chains parallel to the b-axis could also
play a role in superconducting YBa$_{2}$Cu$_{3}$O$_{7}$. We analyzed in
detail the 3D density of states around the Fermi level, E$_{F}$, and found
three Van Hove singularities, two below at -230 and -54 meV, respectively,
and a third one above E$_{F}$ at +27 meV. To this end, we have located in
the 3D E {\it vs. }{\bf k }space, the corresponding saddle points that
produce them. Most importantly, we show that the highest contribution to the
nearest below-E$_{F}$ Van Hove singularity comes from chain states. There is
no consensus about the role that the Van Hove singularities play in the
superconductivity of high-T$_{C}$ superconductors (HTSC) \cite{bokbouvier}
but if they play any at all, then our result shows that the b-axis CuO
chains have to be taken into account for a proper description of
superconductivity in YBa$_{2}$Cu$_{3}$O$_{7}$. Andersen {\it et al. }\cite
{andersen91} stressed long ago the importance of saddle points very close to
the Fermi level. More generally, the composition of the Density of States
(DOS) in the vicinity of the Fermi energy, indicates that a 2D description
of YBa$_{2}$Cu$_{3}$O$_{7}$ could turn out to be insufficient. Pickett {\it %
et al.} \cite{pickett90} emphasized that the c-axis dispersion will lead to
apparent broadening of the Fermi surface in experiments interpreted in terms
of a 2D electronic structure. Also Schuller {\it et al. }found in an early
series of experiments \cite{schuller} that the chains play an important role
in YBa$_{2}$Cu$_{3}$O$_{7-\partial }$ .

The rest of the paper is organized as follows. In section 2, we describe in
detail the three Van Hove singularities that we found. In section 3, we
analyze their composition and the one around E$_{F.}$. Section 4 is devoted
to our conclusions.

\section{\protect\bigskip The Van Hove Singularities}

We have reproduced the band structure calculations of DeWeert {\it et al. } 
\cite{deweert89} for YBa$_{2}$Cu$_{3}$O$_{7}$. We use their tight-binding
version. A certain precision is lost but the essential idea will not change
at all. Since these authors have carefully fitted their {\it ab initio}
bands near the Fermi level, the loss of precision is diminished. We gain, in
exchange, a very quick way of reproducing parts of the energy landscape to
analyze in detail. We have calculated an important number of them to find
the location of the saddle points that produce the Van Hove singularities.

In Fig. 1, we show the Density of States (DOS) about 0.8 eV around the Fermi
level where one expects the most important states that participate in
building the condensate, to lie. The origin is at E$_{F}$. In this range of
energy, we found three Van Hove singularities which we have labelled vHs$_{1}
$, vHs$_{2}$ and VHs$_{3}$. We find VHs$_{1}$ at -230 meV. It is located at 
{\bf k}$_{1}$=(0.42 $\frac{\pi }{a}$, 0.13 $\frac{\pi }{b}$, 0). VHs$_{2}$
is at -54 meV at the high-symmetry point {\bf Y=k}$_{2}${\bf =}(0, $\frac{%
\pi }{b}$, 0). Finally, we found VHs$_{3}$ above E$_{F}$, at +27 meV. This
one is located at {\bf k}$_{3}$=(0, $\frac{\pi }{2b}$, 0). The axes are
labelled {\it a, b, c} as it is conventional.

Notice the high peak at about -200 meV. It comes from a maximum in the 3D E 
{\it vs. }{\bf k }space. None of the singularities appears to be at an
important maximum in the DOS. Only vHs$_{3}$ is located at a small maximum.
One of the ideas behind the Van Hove scenario formulation is that the DOS
would be incremented in an important way by the presence of a singularity in
the vicinity of E$_{F}$ . The 3D DOS does not seem to support this idea.

We found the Van Hove singularities by looking at the energy landscape in
the first Brillouin zone (FBZ) for energies above and below E$_{F}$. We
identified the saddle point that produces each Van Hove singularity from 3D
color plots of the energy landscape (not reproduced in this paper). To check
the saddle point, we calculated the band structure in different paths
passing through it and confirmed that we had a maximum in some direction and
a minimum in the perpendicular one.

In Fig. 2a, we locate VHs$_{1}$. The upper part shows the band structure
from ($\frac{7\pi }{20a}$, 0, 0) to ($\frac{\pi }{2a},\frac{13\pi }{50b},$
0). We indicate the band at the point it has a minimum. In the lower part,
the band structure appears from (0, $\frac{\pi }{4b}$, 0) to ($\frac{17\pi }{%
20a}$, 0, 0). The arrow indicates the place where the maximum appears. The
two paths cross at {\bf k}$_{1}$. In 3D, a saddle point is formed. The
energy at which the saddle point is formed is -230 meV.

Fig.2b, we repeat the procedure for VHs$_{2}$. It is the
nearest-to-the-Fermi-level Van Hove singularity below E$_{F}$. It is located
at -54 meV. It is quite near E$_{F}.$ These states are occupied in the
normal state and should participate in the formation of the superconducting
condensate. For that reason this singularity might play a role \cite
{andersen91}.

In the upper part of Fig. 2b, we can see that this band has a minimum in the
($\frac{-\pi }{a},\frac{\pi }{b},0$) to ($\frac{\pi }{a},\frac{\pi }{b},$0)
direction and a maximum in the perpendicular one from (0, $\frac{\pi }{2b},0$%
) to (0, $\frac{3\pi }{2b},0$). The saddle point is at the high symmetry
point {\bf Y=k}$_{2}$=(0, $\frac{\pi }{b}$, 0). In 3D it forms a saddle
point. The corresponding energy is -54 meV.

For the case of VHs$_{3}$, we have produced Fig. 2c. In this case, we show,
in the upper part of the figure, a path from the center of the FBZ, $\Gamma $%
{\bf =(}0,0,0) to (0, $\frac{\pi }{b},$ 0). Here again the arrow points to
the maximum that identifies the saddle point in this case. The lower part
illustrates the minimum. The paths cross at {\bf k}$_{3}$=(0, $\frac{\pi }{2b%
},$ 0). The corresponding energy is at +27 meV. It is located just above the
Fermi level and it is possible that some electrons are promoted here in the
superconducting state since YBa$_{2}$Cu$_{3}$O$_{7}$ has a critical
temperature, T$_{c}$, of about 100K
Summarizing the results of this section, we found three Van Hove
singularities in the vicinity of E$_{F}$. Two of them are very close to it.
One below (vHs$_{2}$ at -54 meV) and one above (vHs$_{3}$ at 27 meV). These
could play a role since the states in the first should participate in the
formation of the condensate and some electrons could be promoted to the one
next section we will study the composition of these singularities.

\section{Analysis of the Composition.}

A very interesting point to look at is the composition, {\it i.e.,} the
states that contribute to the DOS at each Van Hove singularity. We show this
result in Fig. 3 where we draw explicitly the contribution that we get from
the different states around the Fermi level. The energy at which the Van
Hove singularities appear and the Fermi level, E$_{F}$, are shown. We
present three set of data. The scales are the same. The upper part is
devoted to the contribution coming from the b-axis CuO chains composed by
the atoms labelled Cu(1) and O(1), as it is conventional. In the middle
part, we show the contributions coming from the CuO$_{2}$ planes. These
atoms are labelled Cu(2), O(2) and O(3). In the bottom part of Fig. 3, the
contributions from the apical oxygen O(4) and from the Y atom are shown.
These contributions are calculated here per atom, not per unit cell. We will
do that below. Around E$_{F}$, the Y-contribution is negligeable.

The contribution to the DOS from atoms belonging to the chain, Cu(1) and
O(1) are roughly of the same size, in the interval around E$_{F}$ that we
are considering, although the one from Cu(1) seems to be always slightly
higher than the one coming from O(1) (see upper part). In the middle part of
the figure, the corresponding contribution from the atoms that belong to the
planes, for energies in the interval from about -250 meV to E$_{F},$ differ
strongly. The Cu(2) and O(2) (on the a-axis) atoms, both contribute about
the same and their contribution is comparable to the chain atoms, in this
interval of energy. On the contrary, the states associated to the O(3) atom
(on the b-axis, where the chains lie as well) contribute only about one half
as compared to the former ones. In the lower part of Fig. 3, we can see the
important contribution to the DOS at the energies that we are considering
that comes from the apical atom O(4).

A high peak at about -200 meV is composed mainly of states that come from
O(1) and C(1), the b-chain atoms, and from the O(4) apical atom. At this
energy we find a maximum in the E {\it vs. }{\bf k }space not a VHs. These
states could take part in the formation of the condensate and, more
generally, in the thermodynamics of the superconducting state but they are
not included in a 2D formulation.

Let us concentrate now on the singularities. We see from Fig 3 that at VHs$%
_{1}$(-230 meV) the contributions from chain atoms Cu(1) and O(1), from the
plane atoms Cu(2) and O(2) (on the a-axis), and from the apical O(4) atom
are of about the same size. Only the O(3) atom (on the b-axis) contributes
noticeably less than the others (see middle part of Fig.3). The contribution
to the DOS from the O(4) atom at this energy is also important as we show in
the lower part of the figure.

At VHs$_{2}$ (-54 meV), the total DOS has a small kink as we can see in Fig.
1. This is the nearest to the Fermi energy VHs which is filled with occupied
states, as we have stated before. At VHs$_{2}$ the overwhelming contribution
per atom comes from the chains as it is clear from Fig. 3. Notice that the
contribution from atoms in the plane is of the same order of magnitude than
the one from O(4). If the VHs are of importance in HTSC, it might be
important to include this other contributions as well. We recall again the
simultaneous simulation of tunneling, specific heat and ultrasonic
attenuation experiments \cite{cucolo} including the planes, the chains and
the c-axis as active scenarios. In that formulation the chains were assumed
to become superconducting by proximity effect. In spite of the fact that the
BCS formulation of the proximity effect does not support the idea that the
chains could become superconducting \cite{carbotteprox}, the idea of
including more scenarios is worth more analysis.

\subsection{Contribution per atom}

At E$_{F}$, each atomic contribution from Cu(1), Cu(2), O(2) and O(3) are
and of about the same order. The O(1) and O(4) atoms contribute about half
the previous one and Y has a very small almost negligeable contribution. We
did not take into account Ba atoms. In the next Table I, we summarize these
results.

\bigskip\ 

\begin{center}
\begin{tabular}{ccccc}
Atom & VHs$_{1}$ & VHs$_{2}$ & E$_{F}$ & VHs$_{3}$ \\ 
Cu(1) & 11 & 32 & 10 & 9 \\ 
Cu(2) & 11 & 6 & 12 & 11 \\ 
O(1) & 10 & 30 & 5 & 4 \\ 
O(2) & 13 & 6 & 12 & 14 \\ 
O(3) & 4 & 5 & 13 & 14 \\ 
O(4) & 11 & 2 & 5 & 4
\end{tabular}
\end{center}

{\it Table I- Contribution to the Density of States from states
associated to different atoms in YBa$_{2}$Cu$_{3}$O$_{7}$. We show rounded 
percentages. The
contribution is per atom not per unit cell. Notice the important one from
the chain atoms to VHs$_{2}$. They do not sum up to 100\%.
For
that we have to take into account that there are several repeated atoms in
the unit cell (two Cu(2), for example, see Table II). Y has a minor
contribution in all cases (not shown) and we did not take into account the
Ba atoms contribution which is negligeable at this energies.}

\bigskip

\subsection{The total contribution in the unit cell.}

The contributions per atom are interesting in showing that atoms located out
of the plane do contribute to the DOS about the same than the in-plane ones.
The real difference seems to be therefore mainly in the number of planes in
a unit cell. We compare here the respective contributions from the planes,
the b-chains and the O(4) apical atom, per unit cell.

In Table II, we compare the contribution of the plane states,{\it \ i.e., }%
the states coming from the O(3), Cu(2), and O(2) atoms, counted twice, the
b-axis chain states, Cu(1)-O(1), (counted once) and the apical O(4)
(counted twice). These are 10 of the 13 atoms constituting the unit cell.
The contribution of states coming from Y and from the two Ba atoms are
negligeable at these energies. We stress that Table II compares the
percentage contribution per unit cell.
\vskip0.5cm
\[
\begin{tabular}{ccccc}
Subgroup & VHs$_{1}$ & VHs$_{2}$ & E$_{F}$ & vHs$_{3}$ \\ 
Planes & 55 & 32 & 74 & 78 \\ 
Chain & 21 & 62 & 15 & 13 \\ 
O(4)$_{1}$ & 23 & 5 & 10 & 8
\end{tabular}
\]

{\it Table II- The contribution percentages (rounded) from the
planes, the chains and the O(4) states to the Density of States (DOS) per
unit cell (not per atom). Here the sum should be 100\% if all the atoms were
included.}
\vskip0.5cm

From this table, we can conclude that the contribution per unit cell of the
planes is very important. Therefore, a 2D description of HTSC takes into
account the most important quantitative contributions. This is in agreement
with the relative success of the Van Hove scenario formulation. But from the
same table, we see that the contribution to the DOS from the chains is not
only important but also that the nearest-to-E$_{F}$ occupied Van Hove
singularity is populated by states coming{\bf \ mainly} from the chains.
These states should take part in the formation of the condensate when in the
superconducting state and it will not be surprising that in YBa$_{2}$Cu$_{3}$%
O$_{7}$ at least part of the condensate lies in the chains as Dow and
Blackstead have postulated for superconducting PrBa$_{2}$Cu$_{3}$O$_{7\text{ 
}}$\cite{dow99}. By not including the chains one may lose certain effects as
the simulation of their own experiments by Cucolo {\it et al.}
\cite{cucolo} seems to
prove. The conclusion is that the chains cannot be neglected in a proper
description of YBa$_{2}$Cu$_{3}$O$_{7}$.

{\small \ }

\section{Conclusions}

We have reproduced the band structure calculations by DeWeert{\it \ et al.
}%
\cite{deweert89} 
and have analyzed in detail the energy landscape in the
First Brillouin Zone
for energies around the Fermi level. We have found three Van Hove
Singularities nearby. Two below and one above at energies -230, -54 and +27
meV, respectively. We have label them VHs$_{1}$, VHs$_{2}$ and VHs$_{3}$.
They are located at {\bf k}$_{1}${\bf =}(0.42$\frac{\pi }{a}$, 0.13$\frac{%
\pi }{b}$, 0), {\bf k}$_{2}$=(0,$\frac{\pi }{b}$,0) and {\bf k}$_{3}$=(0,$%
\frac{\pi }{2b}$,0) where the notation is conventional.

In an energy interval of about 100 meV around E$_{F}$, the contribution per
atom of the Cu(1), Cu(2), O(1), O(2), O(3) and O(4) are roughly equal. See
Fig. 3 and Table I. When we take them per unit cell, the planes appear to
contribute in a very important way. This fact shows that from the
quantitative point of view a 2D formulation makes sense. This is in
agreement with the relative success of the Van Hove scenario. But the
contribution of the out-of-plane atoms is very relevant and qualitative
effect are lost when they are not included. The simulation by Cucolo {\it et
al. } \cite{cucolo} of their own experiments seems to illustrate this
point. In VHs$_{2}$,
the nearest-to-E$_{F}$ occupied Van Hove singularity, we find the
contribution from the chains more important than the one coming from the
planes. Also the contribution from the apical O(4) is not at all negligeable
within an interval of energy around E$_{F}$ of the order of 10 times the
gap. At -200 meV, we find a maximum in the E {\it vs. }{\bf k }space. At
this energy, the contribution per atom from the O(4) apical atom is the most
important.

Finally, we conclude, first, that {\bf if }the Van Hove singularities are
relevant to superconductivity in the YBa$_{2}$Cu$_{3}$O$_{7}$, the chains
have to be included. In a more general way, we conclude, from the
composition of the Density of States around E$_{F},$ that though a 2D
description of YBa$_{2}$Cu$_{3}$O$_{7}$ takes into account most of the
weight in the DOS around the Fermi level, it might be insufficient to
account for superconductivity. A 3D scenario seems compulsory to account for
the details. In particular, it would not be surprising that part of the
condensate lies in the chains also in YBaCuO as it was argued to be for
PrBaCuO very recently.

\bigskip
* {\it Permanent address:} Universidad Aut\'onoma Metropolitana, Unidad
Azcapotzalco, Departamento de Ciencias B\'asicas. M\'exico D.F.

\newpage

\begin{figure}[tbp]
\caption{The total density of states (DOS) in the vecinity of the Fermi
energy, E$_F$. We found three Van Hove singularities. One at -230 meV,
another at -54 meV and one just above E$_F$, at +27 meV. The origin is at
E$_F$.}
\label{fig1}
\end{figure}

\begin{figure}[tbp]
\caption{We have located the Van Hove singularities studying
directly the 3D ({\bf k$_z=0$})
energy landscape and finding the saddle points that produce them. Here we
merely show two paths in the FBZ perpendicular to each other. We see that
in one direction E({\bf k}) has a minimum and in the other it has a
maximum. In 3D a saddle point emerges. The upper part (a) is devoted to
VHs$_1$, the middle part (b) to VHs$_2$ and the lower part (c) to
VHs$_3$.}
\label{fig2}
\end{figure}

\begin{figure}[tbp]
\caption{In this set of figures we show the composition of the density of
states (DOS) in an interval of 800 meV around the Fermi level. We find
important contributions at the energies where a Van Hove singularity lies,
from states that do not come from the CuO$_2$ planes although these
contribute the most at E$_F$ to the DOS. See text for more details.}
\label{fig3}
\end{figure}

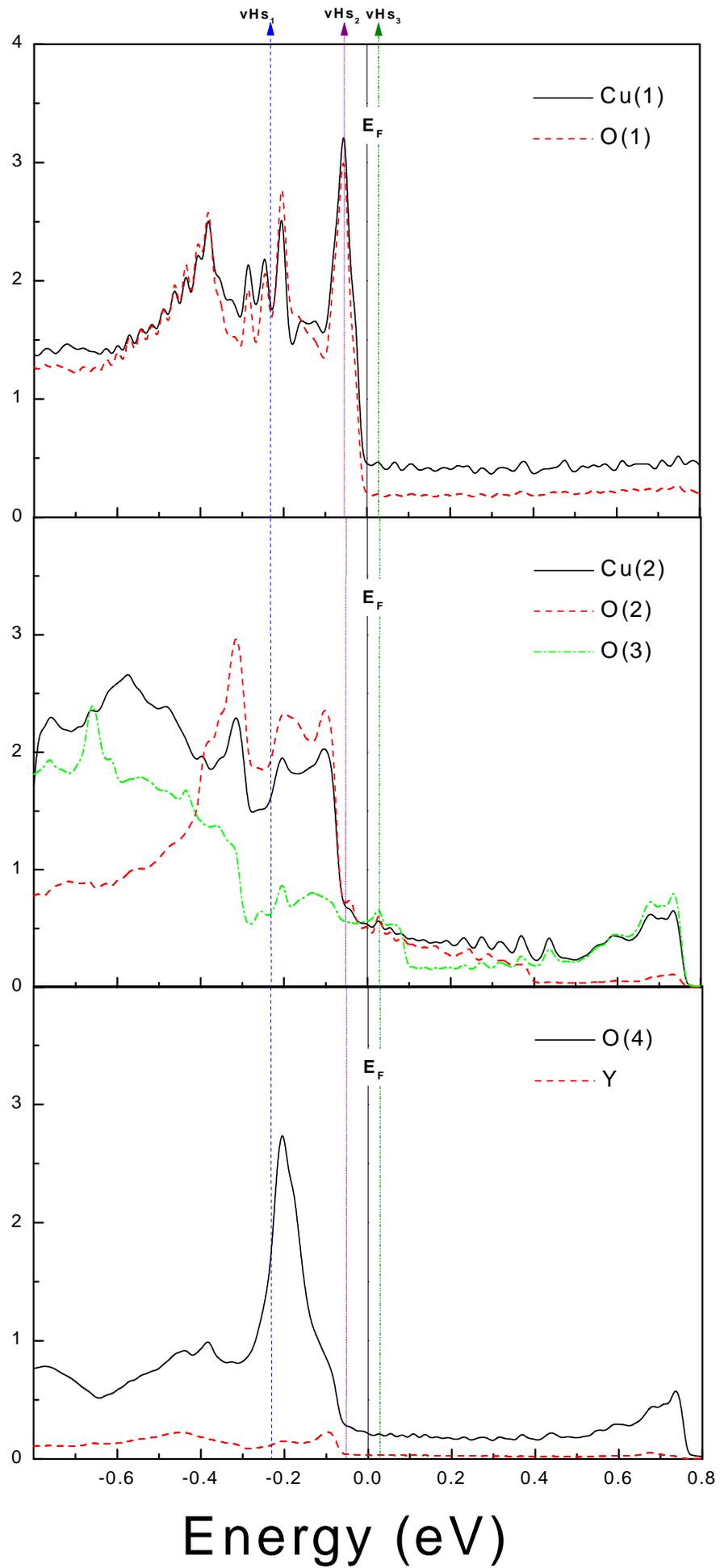
\end{document}